\documentclass[a4paper,12pt]{article}
\usepackage{fullpage}

\usepackage{framed,multirow}

\usepackage{amssymb}
\usepackage{amsmath}
\usepackage{latexsym}
\usepackage{algorithm}
\usepackage{ulem}
\usepackage{multirow}

\usepackage{comment}

\usepackage{authblk}
\usepackage{siunitx}

\usepackage{booktabs}
\usepackage{subfig}
\usepackage{gensymb}
\usepackage{natbib}

\usepackage{algorithmicx}
\usepackage{algpseudocode}

\usepackage[switch]{lineno}

\usepackage{xcolor}
\usepackage{graphicx}

\begin{document}

\title{Improving Lesion Volume Measurements on Digital Mammograms}



\author[a,b]{Nikita Moriakov}
\author[c]{Jim Peters}
\author[b]{Ritse Mann}
\author[b]{Nico Karssemeijer}
\author[c]{Jos van Dijck}
\author[c]{Mireille Broeders}
\author[a,b]{Jonas Teuwen}


\affil[a]{Department of Radiation Oncology, Netherlands Cancer Institute, the Netherlands}
\affil[b]{Department of Medical Imaging, Radboud University Medical Center, the Netherlands}
\affil[c]{Department for Health Evidence, Radboud University Medical Center, the Netherlands}

\maketitle

\begin{abstract}
Lesion volume is an important predictor for prognosis in breast cancer. However, it is currently impossible to compute lesion volumes accurately from digital mammography data, which is the most popular and readily available imaging modality for breast cancer. We make a step towards a more accurate lesion volume measurement on digital mammograms by developing a model that allows to estimate lesion volumes on processed mammograms, which are the images routinely used by radiologists in clinical practice as well as in breast cancer screening and are available in medical centers. Processed mammograms are obtained from raw mammograms, which are the X-ray data coming directly from the scanner, by applying certain vendor-specific non-linear transformations. At the core of our volume estimation method is a physics-based algorithm for measuring lesion volumes on raw mammograms. We subsequently extend this algorithm to processed mammograms via a deep learning image-to-image translation model that produces synthetic raw mammograms from processed mammograms in a multi-vendor setting. We assess the reliability and validity of our method using a dataset of 1778 mammograms with an annotated mass. Firstly, we investigate the correlations between lesion volumes computed from mediolateral oblique and craniocaudal views, with a resulting Pearson correlation of 0.93 [95\% confidence interval (CI) 0.92 - 0.93]. Secondly, we compare the resulting lesion volumes from true and synthetic raw data, with a resulting Pearson correlation of 0.998 [95\%CI 0.998 - 0.998] . Finally, for a subset of 100 mammograms with a malign mass and concurrent MRI examination available, we analyze the agreement between lesion volume on mammography and MRI, resulting in an intraclass correlation coefficient of 0.81 [95\%CI 0.73 - 0.87] for consistency and 0.78 [95\%CI 0.66 - 0.86] for absolute agreement. In conclusion, we developed an algorithm to measure mammographic lesion volume that reached excellent reliability and good validity, when using MRI as ground truth. The algorithm may play a role in lesion characterization and breast cancer prognostication on mammograms. 
\end{abstract}

\section{Introduction}
Breast cancer is the leading cause of cancer-related deaths among women, and it has been shown that breast cancer screening with digital mammography reduces breast cancer mortality \citep{Plevritis2018, zielonke2020evidence}. Its effectiveness and relatively low cost make digital mammography the most popular and readily available imaging modality for breast cancer. 

Prognosis of eventual breast cancer is, among other factors, strongly influenced by the number of tumor cells present, i.e., the volume of the tumor \citep{budczies2015}. A one-dimensional measurement of tumour size is an important predictor of breast cancer prognosis, but knowing the three-dimensional tumour volume may characterize lesions better and lead to a better assessment of breast cancer prognosis \citep{phung2019prognostic,hwang2018prognostic,bozek2014use}. Since mammography is a two-dimensional imaging modality unlike MRI and Breast CT, determining lesion volume directly from lesion segmentation is impossible.

One can make certain simplifying assumptions, for instance, by agreeing that the tumor has a roughly elliptical shape where the major axis of the three-dimensional ellipsoid is parallel to the imaging plane.\citep{peer1993age,heuser1979growth} For such a simplified model, it would suffice to perform a principle component analysis of the segmentation and find the major and the minor axis of the segmentation. This assumption, however, is clearly violated for spiculated lesions.

An improvement could be sought in using the exact area of a segmentation, but this would still require assumptions regarding the three-dimensional shape of a lesion. By using the raw pixel intensities on a mammogram it is possible to obtain estimates of the volume of dense tissue (and, as we propose, the volume of a lesion). Here, the problem of computing the lesion volume is further complicated by the fact that the mammograms that are routinely used by radiologists in clinical practice are the processed mammograms. Processed mammograms are obtained from raw mammograms, which record the ray intensities coming directly from the detector. Processing is performed in the software provided by the vendors, and full details are not disclosed. Such processing is necessary, since the raw data coming from the scanner has an intensity range which makes it unsuitable for visual inspection by a radiologist. During the processing step, the dynamic range is compressed, and the contrast is enhanced to improve the separation between dense and fatty tissues. As an example, one can consider the transformation
\begin{equation}
\textrm{raw} \mapsto \left(C - \mathrm{log}\left( \textrm{raw + 1}\right) \right)^{\gamma}
\end{equation}
with $C := \mathrm{sup} \ \mathrm{log}\left( \textrm{raw + 1}\right) \geq 0$ for some $\gamma \geq 1$, where the dynamic range is compressed by applying the logarithm, the subtraction operation inverts the intensities and the gamma transformation serves to increase the contrast.

In this work, we make a step towards a more accurate breast lesion volume measurement on mammograms using a combination of a physics-based algorithm and deep learning. At the core of our volume estimation approach is a physics-based algorithm for estimating lesion volumes on raw mammograms, which is inspired by a well-known technique for breast density estimation from \cite{vengeland2006}. We subsequently extend our algorithm to processed mammograms via a deep learning image-to-image translation model that produces synthetic raw mammograms from processed mammograms in a multi-vendor setting. We assess the reliability of the method by comparing measurements between the mediolateral oblique (MLO) and cradiocaudal (CC) views, as well as between raw and processed mammograms, and we demonstrate the validity of the mammographic tumour volume measurements using tumor volumes on breast MRI as ground truth.

The article is structured as follows. We describe the data set that was used to train and evaluate our algorithms in Section \ref{s.data}. The methods are described in Section \ref{s.methods}. We show in Section \ref{s.raw} that given a raw mammogram it is possible to obtain a reasonable approximation for the true lesion volume provided that a `lesion-free' version of the mammogram can be guessed, where all lesion tissue is replaced by fat, and we propose a simple algorithm based on rejection sampling of image intensities around the lesion area for this inpainting task. This tumor measurement algorithm is then extended from raw to processed mammograms in Section \ref{s.proc} via image-to-image translation based on pix2pix model \citep{isola2017}. Since we use MRI tumor volumes as ground truth, in Section \ref{s.mriseg} we describe the tumor segmentation model that we use to segment tumors in breast MRI volumes. The results are presented in Section \ref{s.results}. In Section \ref{s.resraw} we investigate the correlation between the tumor volumes obtained on the MLO and CC views, and we stratify these correlations into different breast densities. In Section \ref{s.resproc} we investigate the correlation between the volumes obtained on raw mammograms and the volumes obtained on the corresponding processed mammograms. Finally, in Section \ref{s.resmri} we investigate the agreement between tumor volumes obtained with our algorithm from digital mammograms and the tumor volumes computed from breast MRI.

\section{Data}
\label{s.data}
To develop and validate our methods we relied on data from two common breast imaging modalities: digital mammograms and breast MRI. Digital mammograms were used to train the image-to-image translation model, evaluate the correlations between processed and raw volume measurements as well as the correlations between volumes obtained from different views to assess the reliability of our algorithm. To investigate the validity of the algorithm, breast MRI data was used as the ground truth, where tumor volumes measured on breast MRI were compared with the volumes given by our algorithm for mammograms. For this task, a tumor segmentation model was first trained on a small breast MRI dataset for which ground truth tumor segmentations were provided and assessed by a radiologist, and subsequently this model was used to segment tumors in new volumes.

\subsection{Digital mammograms}
\label{s.mgdata}
For training the deep learning image-to-image translation model, we randomly extracted 6427 pairs of processed and raw mammograms from 1589 patients from our institutional archive. No additional selection criteria were used, thus some images were of healthy patients and some images contained abnormalities. Breast outline was manually segmented in 824 of these images for training of the segmentation model.

\begin{figure*}
\centering
\includegraphics[width=0.9 \textwidth]{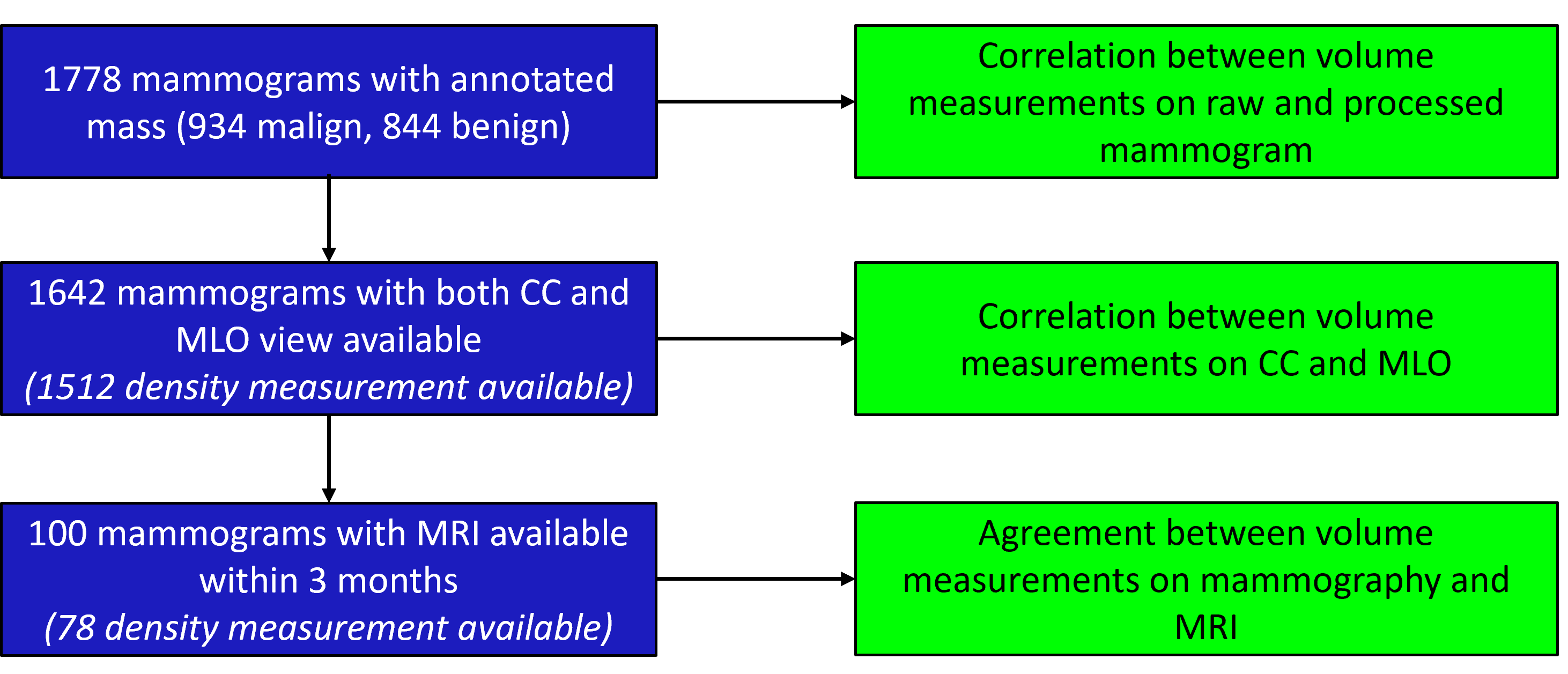}
\caption{Overview of mammograms and MRI images that were used to assess the reliability and validity of lesion volume measurement on mammograms} 
\label{fig.flowchart}
\end{figure*}

Figure \ref{fig.flowchart} shows an overview for the data that has been used to assess the reliability and validity of the developed algorithm. In order to evaluate the volume estimation tool on processed vs. raw mammograms, we extracted all digital mammograms from our archive that met the following criteria:
(1) both the processed and raw mammogram are available;
(2) there is a mass with contour segmentation visible;
(3) the mammogram was not one of the 6427 pairs of processed and raw mammograms that was already used to train the image-to-image translation model.

This selection process resulted in 1778 cases (934 malign and 844 benign masses), with roughly $19 \%$ of the cases acquired on a Mammomat Inspiration (Siemens, Germany), $51 \%$ taken on Senographe DS (General Electric,  USA) and $30 \%$ taken on Senographe 2000D (General Electric,  USA). For 1654 of these cases, both the MLO and CC view were available, allowing for a comparison between views. To assess whether breast density could affect the reliability of our method, we used the volumetric breast density percentage and categories from Volpara\textsuperscript{\tiny\textregistered} Density\textsuperscript{\tiny TM} software, which was available for 1512 cases. All images were re-sampled to the resolution of 200 microns for the subsequent steps.

\subsection{Breast MRI}
\label{s.mrdata}
To obtain tumor volumes on breast MRI, a deep learning tumor segmentation model was trained on a dataset of dynamic contrast-enhanced breast MRI scans of women with tumors that were examined in our institution as a part of breast cancer surveillance program for women at increased risk of breast cancer from January 1, 2003 to January 1, 2014 \citep{vreemann2018} Imaging protocols varied over time and examinations were performed either on 1.5- or a 3-T scanner. The volumes were resampled to isotropic resolution of $0.71$ mm. The dataset was split into 149 subtraction volumes for training and 11 for validation of the model and early stopping of the training when the optimal validation loss is reached. 

To validate our volume estimation algorithm, matching dynamic contrast-enhanced breast MRI scans that were acquired no more than 3 months apart from the corresponding digital mammogram were extracted from the institutional archive. Similar to the training dataset, pre-contrast and post-contrast volumes were registered using deformable registration in elastix \citep{klein2010}. Similarly to the train data, the volumes were resampled to isotropic resolution of $0.71$ mm. To ensure that tumors on two modalities can be reliably matched and that tumor volume measurements on Breast MRI can serve as a ground truth, we limited ourselves only to those Breast MRI volumes with a single lesion presenting itself as a well-defined enhanced mass. Furthermore, in this part of the analysis, we only included malign masses, as characterization of malign masses would be the eventual application of our volume measurement algorithm. The tumor segmentation algorithm was executed on the selected MRI volumes, the quality of the resulting segmentations was controlled visually and they were corrected if necessary. This selection process resulted in a dataset of 100 breast MR volumes matched to mammograms with reliable tumor volume measurements. The MRI volumes were extracted directly from the segmentation.

For all parts of this study, images were pseudonymized and no images were used of patients who objected against the use of their data for scientific research. The need for informed consent was waived and no approval from a Institutional Review Board was required in accordance with the Dutch Medical Research Involving Human Subjects Act.

\section{Methods}
\label{s.methods}
\subsection{Volume estimation baseline}
\label{s.dmbaseline}
Mammography is a two-dimensional imaging modality, therefore certain simplifying assumptions need to be made in order to estimate tumor volume from the linear measurements alone. We will compare our algorithm to the baseline where either the major/minor principal axis of the tumor segmentation mask are measured, and the volume is computed assuming that the lesion has either spherical, cylindrical or elliptical shape with major principal axis parallel to the imaging plane \citep{tilanus2005, langedijk2018, heuser1979growth, peer1993age}. Given the major and minor principal axis of the tumor segmentation mask $a$ and $b$ respectively, we estimated tumor volume under different shape assumptions as follows:

\begin{equation}
\label{eq.vcylinder}
V_{cylinder} = \frac{1}{2} \cdot \biggl(\frac{2a+b}{6}\biggl)^2
\end{equation}
\begin{equation}
\label{eq.vspheroid}
V_{spheroid} = \frac{\pi a b}{6} \cdot \biggl(\frac{a}{2} + \frac{b}{2}\biggl)
\end{equation}
\begin{equation}
\label{eq.vsphere1}
V_{sphere1} = \frac{4 \pi}{3} \cdot \biggl(\frac{2a+b}{6}\biggl)^3
\end{equation}

As a fourth baseline method to compare our algorithm to, we calculated the exact area $A$ of the lesion, based on the lesion mask and the pixel size, and estimated tumor volume assuming it has a roughly spherical shape as follows:
\begin{equation}
\label{eq.vsphere2}
V_{sphere2} = \frac{2A^{\frac{3}{2}}}{3\sqrt{\pi}}
\end{equation}

\subsection{Volume estimation on raw images}
\label{s.raw}
The technique for volumetric breast density estimation from \cite{vengeland2006}, which we rely on in our algorithm for raw images, has been shown to give breast densities that correlate well with the ground truth measurements on MRI. We briefly summarize the physics considerations here, and refer to the original paper for more details.

\begin{figure}
\centering
\includegraphics[width=0.9 \linewidth]{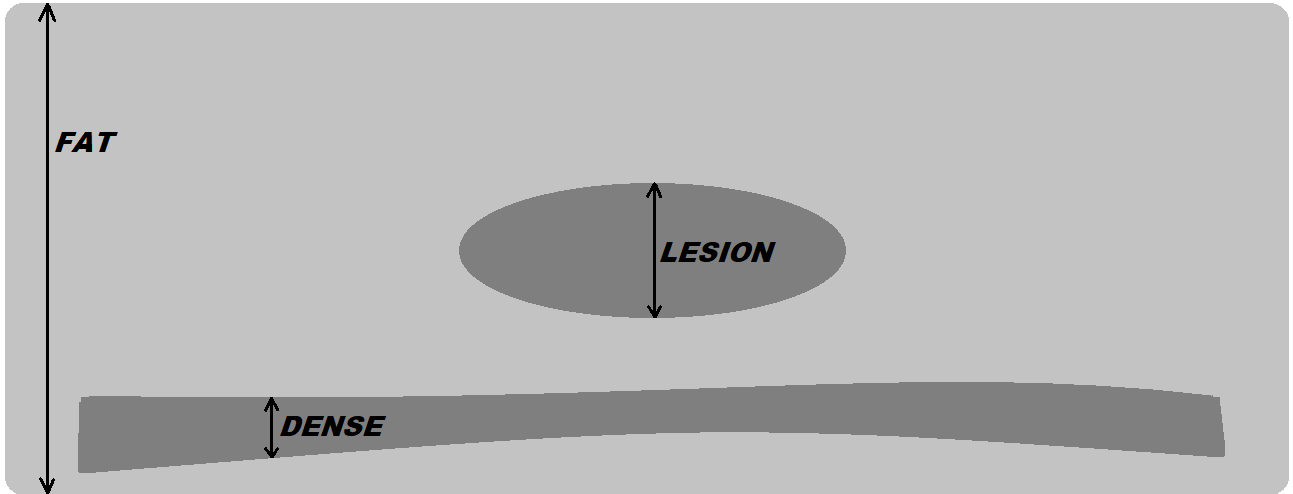}
\caption{Schematic representation of the breast with a lesion.} 
\label{fig.breast}
\end{figure}

For simplicity, we assume that breast is composed of fat, healthy dense tissue and lesion tissue as in Figure \ref{fig.breast}.

To express the pixel intensity $g(r)$ at detector location $r$, we use the following attenuation model for raw mammograms 
\begin{equation}
\label{eq.mammolesion}
\frac{g(r)}{g_0} = \int\limits_{E=0}^{\infty} p(E) e^{-\mu_f(E) h_f(r) - \mu_d(E) (h_d(r) + h_l(r))} dE.   
\end{equation}
Here $p(E)$ is the photon energy spectrum, $\mu_f(E)$, $\mu_d(E)$ are the attenuation coefficients for energy $E$, and $h_f(r)$, $h_d(r)$, $h_l(r)$ are the thicknesses of fatty, dense and lesion tissue respectively for detector location $r$. We assume in the model that the attenuation coefficient of healthy dense tissue equals the attenuation coefficient of lesion tissue. $g_0$ is the ray intensity for an unobstructed pixel, which is often clipped in practice due to e.g. limited dynamic range of the detector. Suppose now that we have access to a `healthy' version of the image $\overline g$, where all lesion tissue is replaced with fatty tissue. Applying similar attenuation model, we get the equation
\begin{equation}
\label{eq.mammonolesion}
\frac{\overline g(r)}{g_0} = \int\limits_{E=0}^{\infty} p(E) e^{-\mu_f(E) (h_f(r) + h_l(r)) - \mu_d(E)h_d(r)} dE.   
\end{equation}
Dividing Eq. \eqref{eq.mammolesion} by Eq. \eqref{eq.mammonolesion}, we conclude that
\begin{equation}
\label{eq.mammoratio}
\frac{g(r)}{\overline g(r)} = \frac{\int\limits_{E=0}^{\infty} p(E) e^{-\mu_f(E) h_f(r) - \mu_d(E) (h_d(r) + h_l(r))} dE} {\int\limits_{E=0}^{\infty} p(E) e^{-\mu_f(E) (h_f(r) + h_l(r)) - \mu_d(E)h_d(r)} dE}.   
\end{equation}
In \cite{vengeland2006}, the authors note that for typical spectra in mammographic imaging such integrals can be well-approximated with exponential functions for some effective attenuation coefficients $\overline{\mu}_{f}$, $\overline{\mu}_{d}$ for fat and dense tissue respectively that depend on acquisition parameters such as the anode and the filter material. We apply a similar approximation to Eq. \eqref{eq.mammoratio}, and after simplifying the fraction we arrive at the identity
\[
\frac{g(r)}{\overline g(r)} = e^{-h_l(r) (\overline{\mu}_{d} - \overline{\mu}_{f})}.
\]
From this identity, we conclude that the lesion thickness $h_l(r)$ at detector location $r$ can be estimated as
\[
h_l(r) = \frac 1 {\overline{\mu}_{f} - \overline{\mu}_{d}} \cdot  \mathrm{ln} \left( \frac{g(r)}{\overline g(r)}\right).
\]
Therefore, the lesion volume $V$ can be estimated as an integral of the lesion thickness $h_l(r)$ over the projection of the lesion, i.e.,
\begin{equation}
\label{eq.volintegral}
V = \frac 1 {\overline{\mu}_{f} - \overline{\mu}_{d}} \int_{r \in \textrm{Lesion}} \mathrm{ln} \left( \frac{g(r)}{\overline g(r)}\right) dr^2
\end{equation}
We conclude that the task of estimating lesion volume can be approached if we have a method for producing `healthy' raw mammograms from mammograms with lesions.

Mathematically speaking, let $g \subset \mathbf R^{n \times m}$ be a raw mammogram of dimensions $n \times m$ and $S \subset [0, n] \times [0, m]$ be a tumor region in $g$. Let $\overline g$ be a `lesion-free' version of $g$ such that $g|_{[0, n] \times [0, m] \setminus S} = \overline g|_{[0, n] \times [0, m] \setminus S}$. The problem of inferring $\overline g$ from $g$ can be viewed as an image inpainting problem, where from a healthy masked image $g|_{[0, n] \times [0, m] \setminus S}$ the goal is to infer $\overline g$. In general, we cannot infer the `true' healthy mammogram with absolute certainty. In dense breasts, for instance, lesion might overlap with regions of healthy dense tissue,  which cannot be distinguished from the lesion based on a single mammogram. At the same time, in fatty breasts the lesion can be completely isolated from dense tissue, which would make the task of inpainting trivial.

Image inpainting have been widely investigated in other contexts, and deep learning-based generative models have been proposed in e.g. \cite{yu2018generative}. For the purposes of this work, however, we developed a simple stochastic inpainting algorithm based on rejection sampling. The algorithm loops over the segmented tumor region $S$, and for each pixel $p \in S$ the algorithm performs rejection sampling of pixel locations from isotropic Gaussian distribution $\mathcal N(p, \mathrm{max}(\frac d 2, 1))$ with $d = \mathrm{dist}(p, \partial S)$ (in mm) restricted to the lesion-free region $[0, n] \times [0, m] \setminus S$. We use adaptive standard deviation in the Gaussian distribution to enforce smooth transition across the segmentation boundary, since in practice the lesions are often slightly over-segmented. We define the output intensity $\overline g[p]$ for pixel $p$ as the average of intensities from the sampled locations. We provide the inpainting algorithm as Algorithm 1 below. 

A complete stand-alone tool for volume estimation on raw mammograms from lesion contours was created in C++, combining both the stochastic inpainting algorithm and the volume integration in Eq. \eqref{eq.volintegral}. Volume measurements are computed separately for MLO and CC views. 
\begin{algorithm}
\caption{Stochastic inpainting algorithm}
\begin{algorithmic}[1]
  \Procedure{Inpaint}{$g, S$}\Comment{Inpaint image $g \in \mathbf R^{n \times m}$ within region $S$}
    \State $\overline g \gets g$ \Comment{Initialize the output image $\overline g$}
    \For{$p \in S$} \Comment{Loop over every pixel in region $S$}
        \State $c \gets 0$ \Comment{Initialize the sample counter with $0$}
        \State $v \gets 0$ \Comment{Initialize the output intensity for pixel p}
        \While{$c < 10$} \Comment{Rejection sampling with 10 samples required}
            \State $d \gets \mathrm{dist}(p, \partial S)$ \Comment{Let $d$ be distance from $p$ to lesion boundary}
            \State $q \gets \mathcal N(p, \mathrm{max}(\frac d 2, 1))$ \Comment{Sample position $q$ from isotropic Gaussian} 
            \If{$q \notin S$} \Comment{Test if sample is outside $S$}
                \State $c \gets c + 1$  \Comment{Increment the counter}
                \State $v \gets v + g[q]$ \Comment{Aggregate the average}
            \EndIf
        \EndWhile
        \State $\overline g[p] \gets v / 10$ \Comment{Define output intensity as average over the samples}
    \EndFor
    \State \textbf{return} $\overline g$ \Comment{Return the inpainted image}
  \EndProcedure
\end{algorithmic}
\end{algorithm}

\subsection{Volume estimation on processed images}
\label{s.proc}
Raw mammograms might be unavailable in practice, since they present no value to the radiologist and thus are not stored in the archives. Therefore, it is necessary to extend the techniques from Section \ref{s.raw} to the case when only processed images are available. We approach this problem via supervised image-to-image translation and develop a `processed-to-raw' model for translating processed mammograms to raw mammograms in a multi-vendor setting. The background intensity level in raw mammograms can vary, sometimes it is clipped and sometimes it is not, which cannot be inferred from the processed version of the image alone, and, furthermore, the information in the background is not used during lesion volume measurements. Thus we first developed a simple breast segmentation model to segment the background as class $0$ and the rest - breast and pectoral muscle mostly - as class $1$, which is then frozen and subsequently used during training and evaluation of the processed-to-raw model. 

We employed the pix2pix model \citep{isola2017} with a `light U-net'-style generator for the image translation task. The generator has depth $8$ and the initial convolution layer consists of 32 filters. As usual, the filter count doubles after each downsampling layer up to the maximum of $32 \times 8 = 256$ filters. Due to memory limitations, we did not use double convolution blocks which are employed by the classical U-net, instead, single convolutions preceded by LeakyReLU non-linearity and followed by instance normalization were utilised similarly to \cite{isola2017}. The tracking of running averages in the instance normalization layers was disabled, which we found to be important for model generalization during validation. For the breast segmentation model we used a similar `light U-net' style architecture, but with depth $7$ instead of $8$ since it is a simpler task.

Processed images, which are provided to the segmentation and translation networks as input after optional augmentations, were transformed as
\begin{equation}
\label{eq.proctransform}
\textrm{processed} \mapsto 1.0 - \textrm{processed} \cdot 0.001,
\end{equation}
where the scaling addresses the fact that processed mammograms are stored as unsigned 12-bit integers, and the subtraction serves to invert the intensity range to make the image more similar to the raw mammogram. The input tensor for the processed-to-raw model was defined as a channel-wise concatenation of the transformed processed image and the breast mask computed by the breast segmentation model.

Raw images, which are the target in image translation, were transformed as
\begin{equation}
\textrm{raw} \mapsto \mathrm{log}\left( \textrm{raw}{ + 1}\right),
\end{equation}
which was done to reduce the complexity of the task for the neural network. The output of the processed-to-raw model and the ground truth raw mammogram were both multiplied by the computed breast mask, after which the pix2pix loss was computed, where the weights of the individual loss components were similar to \cite{isola2017}. The models were implemented in PyTorch 1.10.

The breast segmentation model on digital mammograms was trained for $200$ epochs with batch size $6$ with Adam optimizer \citep{kingma2017adam}. Initial learning rate was set to $2 \cdot 10^{-4}$ and was decreased by $50\%$ every $20$ epochs. During the validation, cross-entropy loss and Dice score were computed. The model with the best loss was subsequently used for training and evaluating the processed-to-raw model. Processed-to-raw model was trained for $50$ epochs with batch size $4$ with Adam optimizer. Initial learning rate was set to $2 \cdot 10^{-4}$ and was decreased by $50\%$ every 10 epochs. The model was evaluated in terms of $L^1$ loss, and the best performing model on validation set was picked to perform the translation for tumor volume measurement. 

\subsection{Tumor volumes on breast MRI}
\label{s.mriseg}
To compute tumor volumes on breast MRI volumes we implemented a 3D U-Net architecture from \cite{nikolov2018}. As input, the model was provided with axial slabs of 21 slices, wherein only the central slice is segmented and the remaining slices serve as context. The corresponding slab thickness provides sufficient field of view, while having complete axial slices should help the model to detect asymmetries between left and right breast. To train the model, we used top-k variant of the cross-entropy loss, so that only the normalized contribution of $5 \%$ pixels with the largest segmentation error was taken into account during backpropagation. Top-k cross-entropy was chosen over regular cross-entropy as a way to mitigate class imbalance.

The tumor segmentation model on breast MRI volumes was trained for 100 epochs with batch size $2$ and Adam optimizer. Initial learning rate was set to $10^{-4}$ and was decreased by $50 \%$ every 20 epochs. Best model was chosen on the basis of Dice coefficient.

\subsection{Statistical analysis of reliability and validity}
\label{s.stats}
Firstly, we assessed the reliability of the processed-to-raw model by comparing tumour volumes obtained from synthetic raw mammograms versus those obtained from the raw mammograms directly in the 1778 mammograms for which both a processed and raw image was available. We did this visually by a scatter plot and by calculating Pearson's correlation coefficients and 95\% confidence intervals (CI's).The reliability of the volume measurement algorithm was also assessed by a scatter plot and Pearson's correlation coefficients and 95\% CI's between volumes computed on MLO and CC views, using the 1654 cases that had both views available. The correlations were also calculated separately for malign masses only, as the intended application of measuring lesion volumes is to characterize malign masses. To understand the behaviour of the model better, we repeated the analyses in different density strata for the 1512 mammograms in which density was measured using Volpara version 1.5.0. We used cut-off points of 4.5\%, 7.5\% and 15.5\% to divide the volumetric breast density in four categories (Volpara Density Grade (VDG) A, B, C and D) that correlate well with the four classes of density defined in the Breast Imaging Reporting and Data Systems (BI-RADS)\cite{highnam2012breast,gubern2014volumetric}. We furthermore report the median relative error and its interquartile range (IQR) between volume measurements on both views, which is defined as the median of $|x - y|/\mathrm{min}(x, y)$ over all pairs $(x, y)$ of measurements from MLO and CC views respectively of the same tumors. 

Secondly, we assess the validity of the algorithm to estimate tumor volume on mammograms by investigating the agreement of the measured volume of 100 malign masses on mammograms with tumor volume on concurrent MRI images, obtained directly from the segmentation model described in Section \ref{s.mriseg}. For mammograms, a single volume measurement for each tumor was obtained by averaging between both views. Here, we calculate intraclass correlation coefficients (ICC) for consistency and absolute agreement, using a two-way mixed effects model. As there is no inter-rater variability for the algorithm, we conservatively predefined an ICC of 0.8 or higher to be indicative of good validity. To assess whether the algorithm improves over simpler methods, we also calculate ICCs after determining tumor volumes on mammograms by Eq. \eqref{eq.vcylinder}, \eqref{eq.vspheroid}, \eqref{eq.vsphere1} and \eqref{eq.vsphere2}. Furthermore, we provide a Bland-Altman plot to visually assess the differences between the two measurements. As this analysis indicated proportional bias (the percentages volume difference between volumes on mammograms and MRI are dependent on the actual volumes), mean differences and limits of agreement are calculated in a regression-based approach.\citep{bland1999measuring} In all measures of correlation and agreement we used log-transformed volumes, given the non-normal distribution of volume measurements. 

\section{Results}
\label{s.results}
\subsection{Breast and tumor segmentation models}
Best-performing breast segmentation model for digital mammograms, that was used as a part of the processed-to-raw image translation model, reached Dice score of $0.96$ on the validation set. Best-performing MRI tumor segmentation model reached Dice score of $0.88$ on the validation set.

\subsection{Reliability of volume measurement between raw and synthetic raw mammograms}
\label{s.resraw}
\begin{figure*}[!ht]
    \centering
    \subfloat[]{\includegraphics[width=0.41\linewidth]{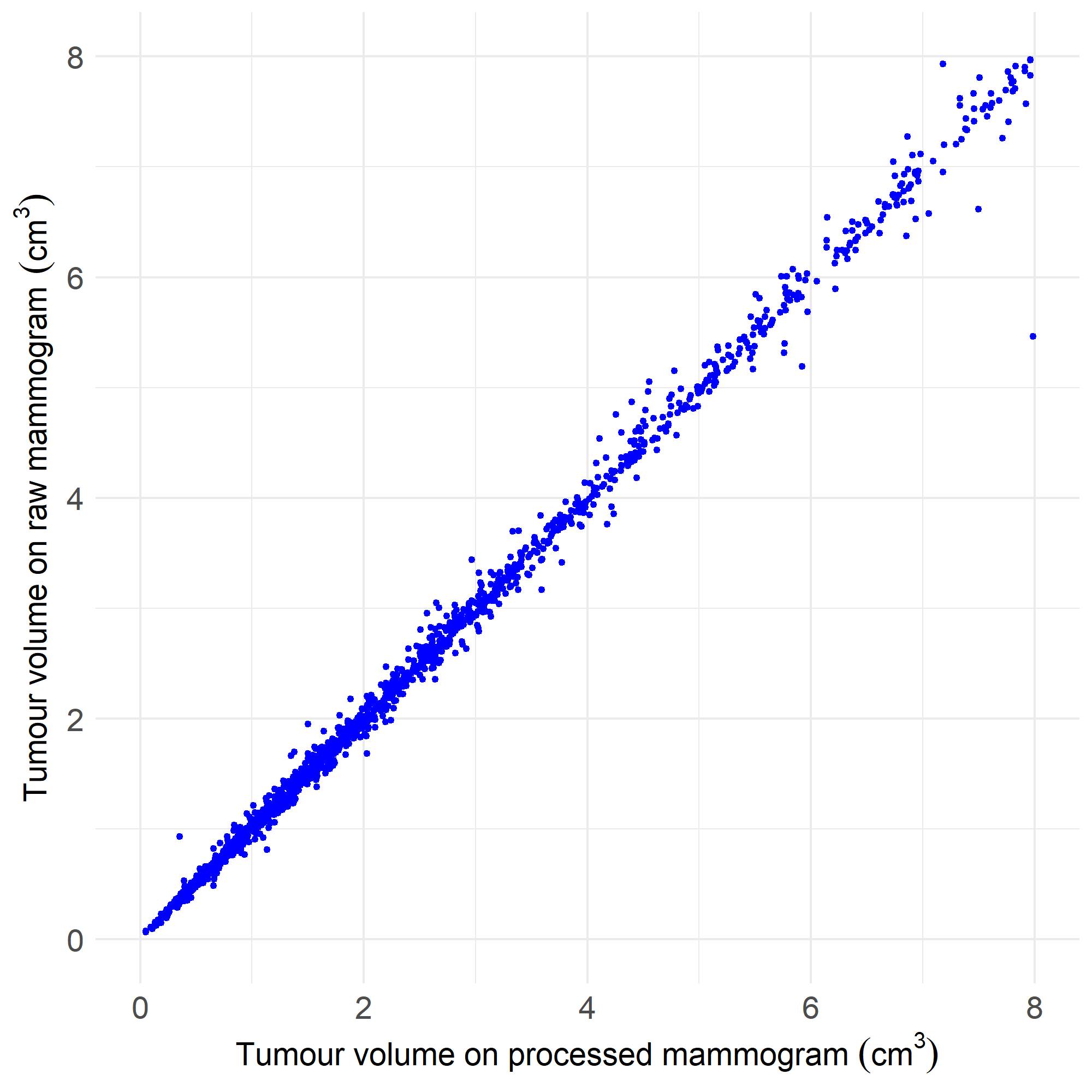}%
    }
    \subfloat[]{\includegraphics[width=0.55\linewidth]{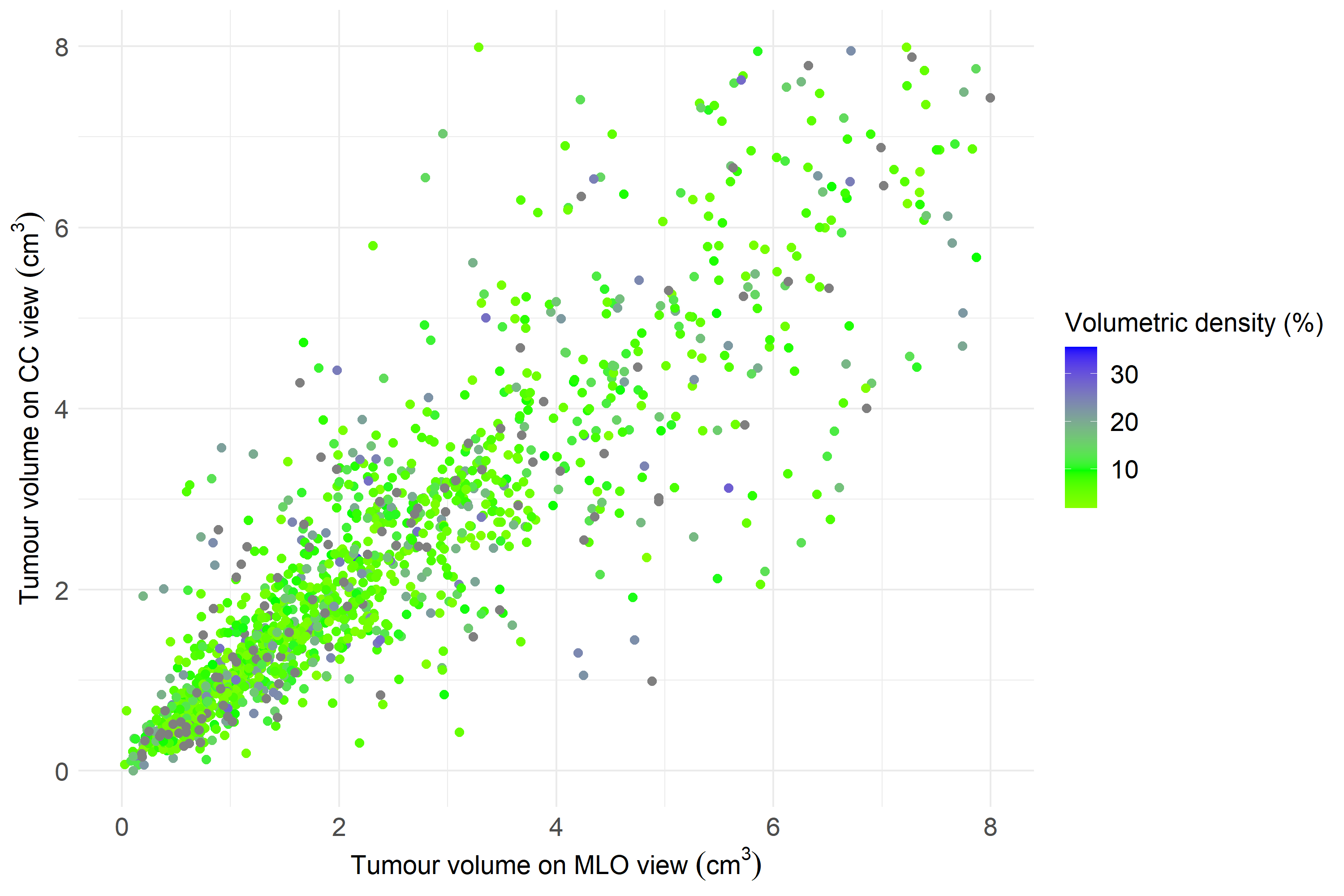}%
    }
    \caption{(a) Scatter plot of tumor volumes measured on raw mammograms directly or after processed-to-raw translation (b) Scatter plot of tumor volumes measured on MLO and CC views}
    \label{fig.mlocc}
\end{figure*}

Figure \ref{fig.mlocc}a shows a scatter plot of tumor volume measurements made on pairs of processed and raw mammograms. The plot axes are limited to a particularly interesting region of volumes $<8$ cm$^3$, which include 88\% of all volume measurements. A visual inspection of volumes outside this range, showed a similar pattern. The volumes were measured on the raw mammogram directly as well as on a synthetic raw mammogram that is generated after applying the processed-to-raw model on a processed mammogram. Based on $1778$ pairs of processed and raw mammograms, the median tumor volume measured by the algorithm was $2.17$ cm$^3$ (IQR $1.11 - 4.59$) when measured on raw mammograms directly and $2.19$ cm$^3$ (IQR $1.10 - 4.50$) when measured on a synthetic raw mammogram. Of the $1778$ lesions that were assessed, 934 were malign. Here, the measured median volumes were $2.80$ cm$^3$ (IQR $1.53 - 5.50$) on processed mammograms and $2.82$ cm$^3$ (IQR $1.56 - 4.98$) on synthetic raw mammograms. Table \ref{tab.rawcorr2} shows the Pearson's correlations between volume measurements on processed and raw mammograms. We observed near perfect correlation for all masses Pearson's r = $0.998$ (95\%CI = $0.998 - 0.998$) and similarly high correlations were observed when only considering malign masses as well as in different VDG strata. We inspected the outliers visually, and concluded that the failures were due to bad breast segmentation for some extremely large breasts, since the number of segmentations used to train our breast segmentation model was somewhat limited.

\begin{table}
\caption{Volume correlation and relative errors between all processed and raw mammograms and stratified by breast density categories}
\centering
\label{tab.rawcorr2}
\begin{tabular}{|c|c|c|c|m{70pt}|}
\hline
Density &  Masses & N cases & Pearson's r (95\%CI) &  Median relative error \% (IQR)\\
\hline
All & All & 1778 & 0.998 \textit{(0.998 - 0.998)} & 1.9 \textit{(0.8 - 4.3)}\\
& malign & 934 & 0.998 \textit{(0.998 - 0.998)} & 1.7 \textit{(0.8 - 3.9)}\\
\hline
A & All & 325 & 0.996 \textit{(0.996 - 0.997)} & 2.5 \textit{(1.2 - 5.0)}\\
& malign & 163 & 0.995 \textit{(0.993 - 0.996)} & 2.0 \textit{(1.2 - 4.9)}\\
\hline
B & All & 460 & 0.999 \textit{(0.999 - 0.999)} & 1.5 \textit{(0.7 - 3.1)}\\
& malign & 234 & 0.999 \textit{(0.999 - 0.999)} & 1.4 \textit{(0.7 - 2.6)}\\
\hline
C & All & 566 & 0.999 \textit{(0.999 - 0.999)} & 1.6 \textit{(0.7 - 3.6)}\\
& malign & 314 & 0.999 \textit{(0.999 - 0.999)} & 1.4 \textit{(0.7 - 2.8)}\\
\hline
D & All & 282 & 0.999 \textit{(0.999 - 0.999)} & 2.0 \textit{(0.7 - 5.0)}\\
& malign & 126 & 0.999 \textit{(0.999 - 0.999)} & 2.0 \textit{(0.7 - 5.5)}\\
\hline
\end{tabular}
\end{table}

\begin{table}
\caption{Volume correlation and relative errors between all MLO and CC views and stratified by breast density categories}
\centering
\label{tab.viewcorr}
\begin{tabular}{|c|c|c|c|m{70pt}|l|}
\hline
Density &  Masses & N cases & Pearson's r (95\%CI) &  Median relative error \% (IQR)\\
\hline
All & All & 1642 & 0.93 \textit{(0.92 - 0.93)} & 20.6 \textit{(8.7 - 43.9)}\\
& malign & 873 & 0.92 \textit{(0.91 - 0.93)} & 19.8 \textit{(8.2 - 42.93)}\\
\hline
A & All & 305 & 0.95 \textit{(0.93 - 0.96)} & 19.0 \textit{(8.3 - 32.3)}\\
& malign & 154 & 0.95 \textit{(0.91 - 0.97)} & 20.2 \textit{(10.4 - 34.0)}\\
\hline
B & All & 429 & 0.92 \textit{(0.91 - 0.93)} & 17.8 \textit{(7.6 - 38.1)}\\
& malign & 219 & 0.91 \textit{(0.88 - 0.92)} & 16.3 \textit{(7.0 - 33.8)}\\
\hline
C & All & 528 & 0.94 \textit{(0.93 - 0.95)} & 22.1 \textit{(8.7 - 47.0)}\\
& malign & 298 & 0.93 \textit{(0.92 - 0.95)} & 22.0 \textit{(9.7 - 47.0)}\\
\hline
D & All & 250 & 0.90 \textit{(0.87 - 0.92)} & 24.1 \textit{(10.9 - 53.8)}\\
& malign & 114 & 0.92 \textit{(0.89 - 0.94)} & 23.8 \textit{(10.4 - 48.3)}\\
\hline
\end{tabular}
\end{table}

\subsection{Reliability of volume measurement between views}
\label{s.resproc}
Based on $1642$ cases with both views available, the median tumor volume was $2.88$ cm$^3$ (IQR $1.54 - 5.85$) when estimated from MLO views and $2.79$ cm$^3$ (IQR $1.48 - 5.29$) on CC views. Figure \ref{fig.mlocc}(b) shows a scatter plot of the volume measurements on both views, where the color bar indicates the volumetric density percentage (on which VDG density categories are based). Table \ref{tab.viewcorr} shows the Pearson correlation's between views. Here, we observe a strong correlation of $0.93$ (95\%CI $0.92 - 0.93$) when considering all masses and a strong correlation of $0.92$ (95\%CI $0.91 - 0.93$) when only malign masses are considered. Furthermore, we did not observe any significant variability across the different VDG density classes. We observed a strong correlation between measurements in all cases, but, in general, the volumes estimated from MLO and CC views can disagree substantially. Visual investigation of the outliers revealed that very often this can be associated with large variability in the area of the tumor segmentation between MLO and CC views. For VDG A/B the median relative error is slightly smaller, because it can partially cancel the effect of this variability by performing the inpainting which allows to isolate the lesion from fatty background.



\subsection{Validity of volume measurement: comparison with breast MRI}
\label{s.resmri}
\begin{figure*}
\centering
\includegraphics[width=\linewidth]{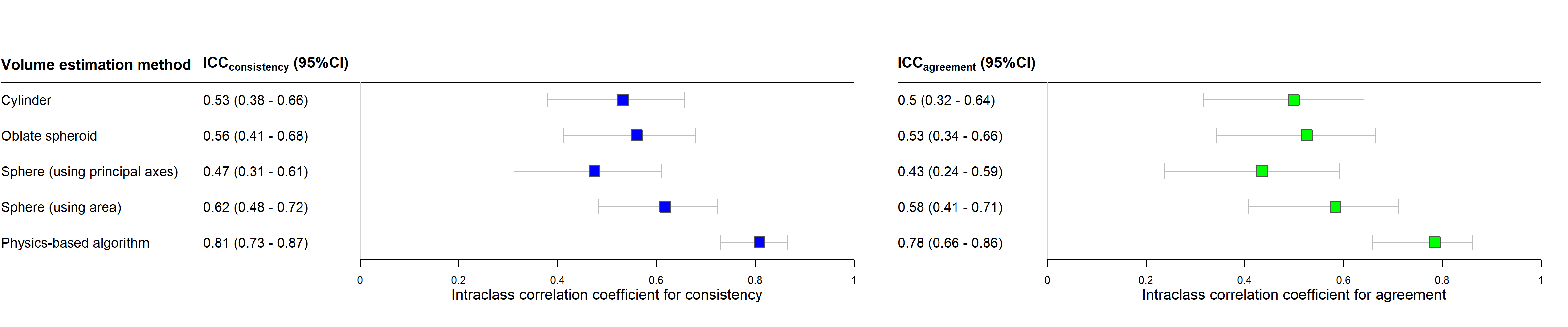}
\caption{Forest plot of ICCs for consistency and agreement between tumor volume measurements on mammograms and MRI} 
\label{fig.forrest}
\end{figure*}

To assess whether the volume measurements obtained on mammograms by the algorithm are measuring the actual tumor volume, we considered MRI as a reference standard and investigated the agreement between measurements on mammograms and MRI for 100 malign cases with both modalities available within 3 months from each other. On mammograms, volumes were computed using our algorithm and for comparison using methods with simplifying assumptions regarding tumor shape and using either the major/minor principal axes or the area of the tumor mask, given in equations \ref{eq.vcylinder} to \ref{eq.vsphere2}. Figure \ref{fig.forrest} is a forest plot showing the ICCs for agreement and consistency for each method, compared to MRI. The algorithm reached ICCs of 0.81 [95\%CI 0.73 - 0.87] for consistency and 0.78 [95\%CI 0.66 - 0.86] for absolute agreement and is an improvement on the other methods and the only method to measure tumor volume with good validity on mammograms. The ICCs and 95\%CIs for agreement and consistency for other methods are displayed in Figure \ref{fig.forrest}. Differences between the validity of the physics-based algorithm and that of other methods were statistically significant, with the exception of the comparison of the ICC for agreement between the algorithm and a measurement method that uses the exact area of the segmentation and \ref{eq.vsphere2} to estimate lesion volume.

\begin{figure*}
\centering
\includegraphics[width=\linewidth]{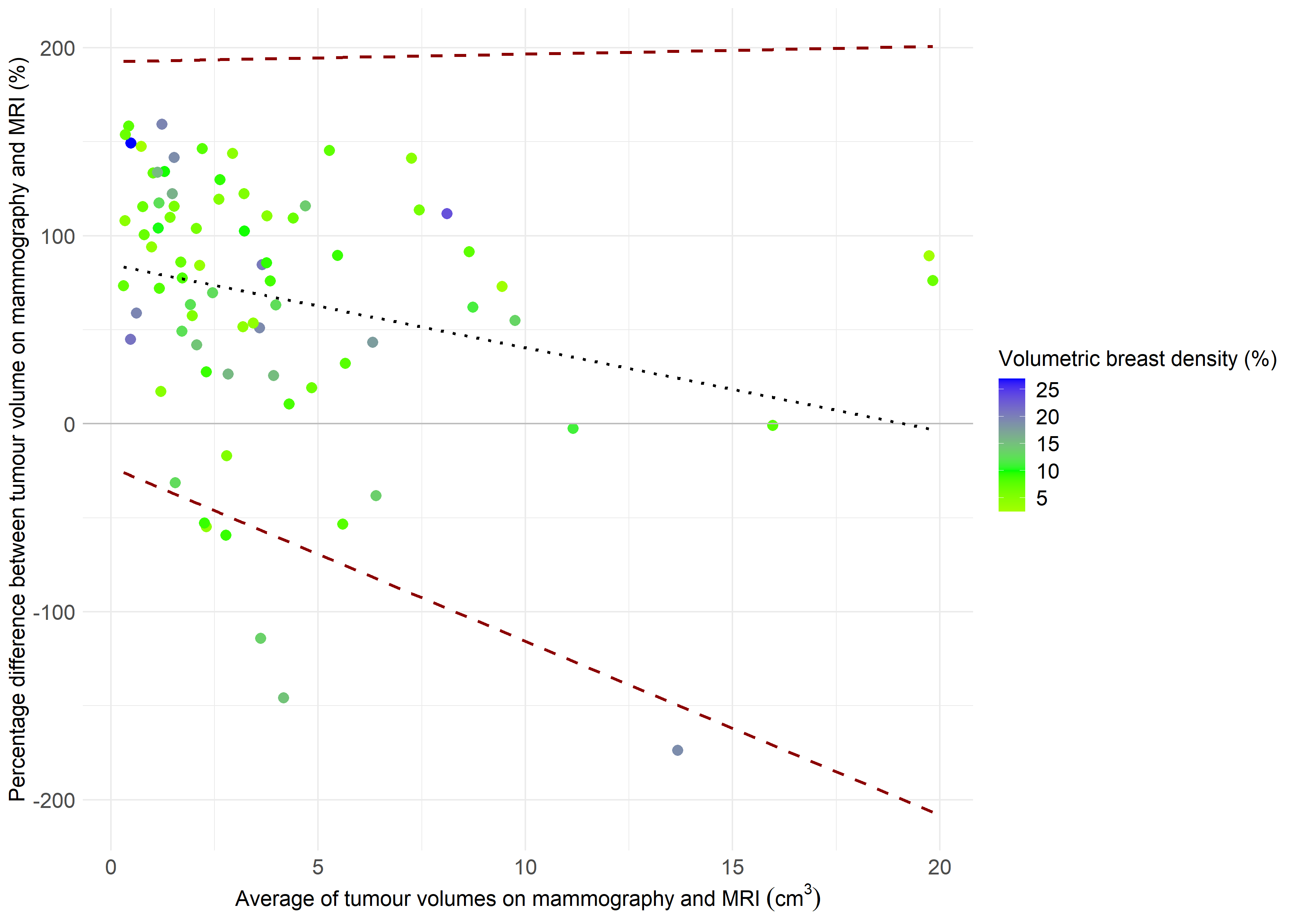}
\caption{Bland Altman plot for the agreement between tumor volume measured on mammograms and MRI} 
\label{fig.BAdens}
\end{figure*}

Using the algorithm, the median tumor volume of the 100 malign masses was 3.09 cm$^3$ (IQR 1.76 - 5.31) when measured with the algorithm on mammograms and 1.66 cm$^3$ (IQR 0.61 - 3.37) when obtained from the segmented mass on MRI. Figure \ref{fig.BAdens} shows the Bland-Altman plot for the agreement of tumor volume measured using our algorithm (averaged over both views) and MRI. The x-axis shows the average of volume measurements on each modality and the y-axis the percentage difference between both. A linear regression of the percentage difference on the average volume yielded a regression intercept of 84.7\% (95\%CI 63.3 - 106.2) and slope of -4.4\%/cm$^3$ (95\%CI -8.3 - -0.6). The mean differences indicate that across the range of volumes the algorithm measuring tumor volume on mammograms overestimates the actual tumor volume, and that the overestimation is largest for small tumors and decreases with increasing tumor size. Despite the high ICC, the limits of agreement (in red in Figure \ref{fig.BAdens}) indicate that for individual cases, substantial over- or underestimation of actual tumor volume may be expected. The plot furthermore does not show any clear relation between the agreement between tumor volume on mammography and different levels of volumetric breast density.

\section{Conclusion and discussion}
\label{s.discussion}
We developed a model that allows to estimate lesion volumes on processed mammograms, which relies on physics-based algorithm for measuring lesion volumes on raw mammograms. We observed excellent correlation for the lesion volumes computed on the original raw data and on the synthetic raw mammograms, and a strong correlation between volumes measured from MLO and CC views respectively. This correlation was strong in all strata of breast density. Furthermore, in a subset of the dataset with cases that had both mammography and a concurrent MRI available, we observed a strong agreement with the MRI measurements. Overall, estimated volumes on mammograms were larger than on MRI and the relative differences were larger for smaller tumors. In individual cases, differences could be substantial. We hypothesize that this may be caused by the fact that the segmentation on MRI more narrowly defines tissue that really contains tumor cells, while on mammography there might be some tissue surrounding the tumor may also be segmented, particularly if the tumor margin does not have a sharp appearance. When visually assessing cases with large relative differences, we also observed examples where a necrotic core was visible on MRI (and not segmented as tumor tissue), while the entire region would be segmented on mammography, thereby overestimating the volume of tumor cells in the region within the tumor margins. 

A limitation of our method related to the segmentation, is the lack of case-specific confidence bounds. We conjecture that this shortcoming can be partially compensated with deep learning algorithms for both probabilistic segmentation \cite{kohl2019hierarchical} and probabilistic inpainting. Another limitation of this study is that conclusions on the validity of volume measurements on mammograms can only be drawn for tumors that presented itself as well-defined malign masses, since this was a selection criterion for MRI data. We do however expect this algorithm to perform well in other settings for the following reasons: 1) The algorithm is based on physics, e.g. it does not `know' what lesion type it is assessing, 2) The clear improvement in agreement with MRI for the algorithm over methods that only use the principal axes or exact area of the segmentation is also not expected to be specific to well-defined masses. In fact, a physics-based approach may offer an advantage over simpler methods in more challenging scenarios. Another limitation of the analysis of validity, is that we did not have enough cases (100) to stratify these analyses on breast density, and thus we cannot provide insight in differential behaviour of the algorithm across density strata. An alternative approach that could theoretically improve agreement between tumor volumes on mammography and MRI, would be to consider tumor volume measurement on mammograms as a prediction task, where MRI measurements are used as ground truth to train a model with information from the mammogram (either pixels from the segmented area directly, e.g. using deep learning, or by recalibrating output from our physics-based model). We have refrained from this approach for three reasons: 1) This would potentially require a larger dataset of concurrent mammograms and MRIs, 2) the physics-based model is more readily interpretable, and it is not obvious what physical differences we are actually accounting for for by recalibrating the model, 3) the physics-based model does not depend on the data and therefore better generalizability may be expected.

In conclusion, we have developed and evaluated an algorithm to assess lesion volume on mammograms. We established that it does so with excellent reliability and good validity and improved over existing methods. A potential clinical application would be to characterize lesions using mammography. In particular when no other information (from either other imaging modalities or pathological assessment) is available, this could give more insight in the prognosis of breast tumors then measuring size alone, which is the current practice in imaging-based staging. Breast cancer screening is a particular interesting area, as mammography is the modality of choice in most screening programs. It offers the additional advantage that for many participants serial mammograms are available, which would allow for the assessment of tumor growth if a lesion is observed during multiple examinations. The exact value of tumor volume as a prognostic marker in different settings does however require more research.

\section{Acknowledgements}
This work is a part of the research conducted by IMAGINE consortium under the Dutch Cancer Society grant, grant number 11835. The IMAGINE consortium is formed by the following researchers: Carla van Gils, Sjoerd Elias, Bas Penning de Vries (Julius Center for Health Sciences and Primary Care, UMC Utrecht, Utrecht, The Netherlands), Ruud Pijnappel (Dutch Expert Center for Screening, Nijmegen, The Netherlands), Esther Lips, Jelle Wesseling, Sandra van den Belt, Merle van Leeuwen (Division of Molecular Pathology, Netherlands Cancer Institute, Amsterdam, The Netherlands), Marja van Oirsouw, Ellen Verschuur (Dutch Breast Cancer Association, Utrecht, The Netherlands), Nico Karssemeijer, Ritse Mann (Department of Medical Imaging, Radboud university medical center, Nijmegen, The Netherlands), Jonas Teuwen, Nikita Moriakov (Department of Radiation Oncology, Netherlands Cancer Institute, Amsterdam, The Netherlands), Jim Peters, Jos van Dijck and Mireille Broeders (Department for Health Evidence, Radboud university medical center, Nijmegen, The Netherlands). The consortium is led by Mireille Broeders.

We would additionally like to thank Riccardo Samperna, who assisted with the data collection for the pairs of mammograms and MRI’s at the time.

\bibliographystyle{abbrvnat}
\bibliography{bibliography}

\end{document}